\RequirePackage[left]{}
\documentclass[
twocolumn%
,amssymb, amsmath,nobibnotes, aps, prl,showpacs,nofootinbib, superscriptaddress]{revtex4}
\usepackage{bm}%
\usepackage{graphicx}
\usepackage[colorlinks=true,linkcolor=blue]{hyperref}%
\expandafter\ifx\csname package@font\endcsname\relax\else
 \expandafter\expandafter
 \expandafter\usepackage
 \expandafter\expandafter
 \expandafter{\csname package@font\endcsname}%
\fi

\newcommand{\japj}{Astrophys. J.}
\newcommand{\jaa}{Astron. Astrophys.}
\newcommand{\jasr}{Adv. Space Res.}
\newcommand{\jprep}{Phys. Rep.}
\newcommand{\jnat}{Nature (London)}
\newcommand{\jarXiv}[1]{arXiv:#1}
\newcommand{\jprl}{Phys. Rev. Letters}

\newcommand{\jproc}[3]{in {\it Proceedings of the #1, #2, #3}}
\newcommand{\jnima}{Nucl. Instrum. Methods Phys. Res. A}

\newcommand{\etal}{{\it et al.}}
\newcommand{\jtbp}{(to be published)}
\newcommand{\jloc}[3]{{\bf #1}, #2 (#3)}
\newcommand{\jref}[4]{#1, \jloc{#2}{#3}{#4}}

\begin{document}

\title{Measurement of the Cosmic Ray $e^{+} + e^{-}$ spectrum from 20 GeV to 1 TeV with the Fermi Large Area Telescope}

\author{A.~A.~Abdo}
\affiliation{National Research Council Research Associate}
\affiliation{Space Science Division, Naval Research Laboratory, Washington, DC 20375}
\author{M.~Ackermann}
\affiliation{W. W. Hansen Experimental Physics Laboratory, Kavli Institute for Particle Astrophysics and Cosmology, Department of Physics and SLAC National Accelerator Laboratory, Stanford University, Stanford, CA 94305}
\author{M.~Ajello}
\affiliation{W. W. Hansen Experimental Physics Laboratory, Kavli Institute for Particle Astrophysics and Cosmology, Department of Physics and SLAC National Accelerator Laboratory, Stanford University, Stanford, CA 94305}
\author{W.~B.~Atwood}
\affiliation{Santa Cruz Institute for Particle Physics, Department of Physics and Department of Astronomy and Astrophysics, University of California at Santa Cruz, Santa Cruz, CA 95064}
\author{M.~Axelsson}
\affiliation{The Oskar Klein Centre for Cosmo Particle Physics, AlbaNova, SE-106 91 Stockholm, Sweden}
\affiliation{Department of Astronomy, Stockholm University, SE-106 91 Stockholm, Sweden}
\author{L.~Baldini}
\affiliation{Istituto Nazionale di Fisica Nucleare, Sezione di Pisa, I-56127 Pisa, Italy}
\author{J.~Ballet}
\affiliation{Laboratoire AIM, CEA-IRFU/CNRS/Universit\'e Paris Diderot, Service d'Astrophysique, CEA Saclay, 91191 Gif sur Yvette, France}
\author{G.~Barbiellini}
\affiliation{Istituto Nazionale di Fisica Nucleare, Sezione di Trieste, I-34127 Trieste, Italy}
\affiliation{Dipartimento di Fisica, Universit\`a di Trieste, I-34127 Trieste, Italy}
\author{D.~Bastieri}
\affiliation{Istituto Nazionale di Fisica Nucleare, Sezione di Padova, I-35131 Padova, Italy}
\affiliation{Dipartimento di Fisica ``G. Galilei", Universit\`a di Padova, I-35131 Padova, Italy}
\author{M.~Battelino}
\affiliation{The Oskar Klein Centre for Cosmo Particle Physics, AlbaNova, SE-106 91 Stockholm, Sweden}
\affiliation{Department of Physics, Royal Institute of Technology (KTH), AlbaNova, SE-106 91 Stockholm, Sweden}
% ...

\collaboration{The Fermi LAT Collaboration}
\noaffiliation
\collaboration{Submitted to Physical Review Letters on March 19, accepted on April 21, published online on May 4}\noaffiliation

\begin{abstract}

Designed as a high-sensitivity gamma-ray observatory, the Fermi Large Area Telescope is also an electron detector with a large acceptance exceeding 2 ${\rm m}^2{\rm sr}$ at 300 GeV.
Building on the gamma-ray analysis, we have developed an efficient electron detection strategy which provides sufficient background rejection for measurement of the steeply-falling electron spectrum up to 1 TeV.
Our high precision data show that the electron spectrum falls with energy as $E^{-3.0}$ and does not exhibit prominent spectral features.
Interpretations in terms of a conventional diffusive model as well as a potential local extra component are briefly discussed.
\end{abstract}
 
\pacs{96.50.sb, 95.35.+d, 95.85.Ry}

\maketitle

{\it Introduction.} -- Accurate measurements of high-energy Cosmic Ray (CR) electrons (we hereafter refer to electrons as a sum of $e^{+}$ and $e^{-}$ unless specified otherwise) provide a unique opportunity to probe the origin and propagation of CRs in the local interstellar medium and constrain
models of the diffuse gamma-ray emission~\cite{SMR2004}.
Prior to 2008, the high-energy electron spectrum was measured by balloon-borne experiments~\cite{kobayashi} and by a single space mission (AMS-01,~\cite{ams1}). 
The measured fluxes differ by factors of $2-3$. 
While these data allowed significant steps forward in understanding CR origin and propagation, constraints on current models remain weak.  
The CR propagation package GALPROP~\cite{galprop}, on the assumption that electrons originate from a distribution of distant sources mainly associated with supernova remnants and pulsars, predicts a featureless spectrum from 10 GeV up to few hundreds of GeV. 
Above that energy, due to the actual stochastic nature of electron sources in space and time, and to the increasing synchrotron and inverse Compton energy losses, the spectral shape may exhibit spatial variations on a scale of a few hundred parsecs.
Nearby sources start contributing significantly to the observed local flux and may induce important deviations from a simple power law spectrum~\cite{kobayashi,aharonian,pohl}.

Recently published results from Pamela~\cite{pamela}, ATIC~\cite{atic}, H.E.S.S.~\cite{hess} and PPB-BETS~\cite{ppbbets} have opened a new phase in the study of high energy CR electrons with a new generation of instruments. 
These four experiments report deviations from the reference model mentioned above.
Pamela measures an increase of positrons with respect to electrons at energies above a few GeV; ATIC and PPB-BETS detect a prominent spectral feature at around 500 GeV in the total electron plus positron spectrum; H.E.S.S. reports significant steepening of the spectrum above 600 GeV.
All these results may indicate the presence of a nearby primary source of electrons and positrons.
The natures of possible sources have been widely discussed, two classes of which stand out: nearby pulsar(s) (~\cite{shen},~\cite{Profumo:2008ms} and references therein) and dark matter annihilation in the Galactic halo, e.g.~\cite{cholis}. 

The Large Area Telescope (LAT) is the main instrument on-board the recently launched Fermi Gamma-Ray Space Telescope mission.
It is conceived as a multi-purpose observatory to survey the variable gamma-ray sky between 20 MeV and 300 GeV, including the largely unexplored energy window above 10 GeV.
It is therefore designed as a low aspect ratio, large area pair conversion telescope to maximize its field of view and effective area.
The LAT angular, energy and timing resolution rely on modern solid state detectors and electronics~\cite{LATpaper}: 
a multi-layer silicon-strip tracker (TKR), interleaved with tungsten converters with a total depth of 1.5 radiation length (X$_{0}$) on-axis;
 a hodoscopic CsI(Tl) calorimeter (CAL), 8.6 X$_{0}$ deep on-axis;
 a segmented anticoincidence plastic scintillator detector (ACD) 
and a flexible, programmable trigger and filter logic for on-board event filtering.

Since electromagnetic (EM) cascades are germane to both electron and photon interactions in matter, the LAT is also by its nature a detector for electrons and positrons.
Its potential for making systematics-limited measurements of CR electrons was recognized during the initial phases of the LAT design~\cite{durban},~\cite{merida}.
In this Letter we describe the technique developed for this measurement and our validations using
ground and flight data, and present the first high statistics CR electron spectrum from 20 GeV to 1 TeV based on the data taken in the first six months of the mission.
A more complete description of the analysis procedure, and new results with increased statistics, energy coverage and a preliminary study of sensitivity to anisotropies in the distribution of arrival directions will be presented in forthcoming papers.

{\it Event selection.} -- An analysis that separates electrons from the dominant CR hadrons is required to fully exploit LAT's large collecting power and long observation time. 
The on-board filter is configured to accept all events that deposit at least 20 GeV in the calorimeter; thus we ensure that the rare high-energy events, including electrons, are available for thorough analysis on the ground.
We developed a dedicated event selection for high energy electrons that provides a large geometry factor with a residual hadron contamination less than $20\%$ at the highest energy (see table~\ref{table:FermiCREdata}).
At any given energy, the geometry factor (GF) is defined as the proportionality factor relating the rate of events passing the selection criteria to the incident flux.
It measures the instrument acceptance, i.e the integral of the effective area over the instrument field of view. 
It is numerically evaluated using a MC data sample of pure electrons.
As for the analysis developed for extracting LAT photon data~\cite{LATpaper}, the electron selection essentially relies on the LAT capability to discriminate EM and hadronic showers based on their longitudinal and lateral development, as measured by both the TKR and CAL detectors.
The background rejection power for photon science is optimized up to 300 GeV.
The electron selection criteria are instead tuned in the multi-100 GeV range, where the much steeper electron spectrum requires an overall hadron rejection power of $1:10^{4}$.

Events considered for the electron analysis are required to fail the ACD vetoes developed to select photon events~\cite{LATpaper}.  
This removes the vast majority of the potential gamma-ray contamination.
In fact, the geometry factor for photons, determined with electron selection cuts, is less than 8\% of that for electrons at 1 TeV. 
The ratio of photon to electron fluxes is negligible at low energies and rises to around 20\% at the highest energy.
This estimate is obtained from a simple, conservative extrapolation to high energies of the EGRET all-sky spectrum at GeV energies using a $E^{-2.10}$ power law~\cite{egretdiffuse}.
The overall gamma contamination in the final electron sample is therefore always less than 2\%.

EM showers start developing in the LAT TKR, while most of the energy is absorbed in the CAL.
The measurement of the lateral shower development is a powerful discriminator between more compact EM showers and wider hadronic showers.
We select on variables that map the distribution of TKR clusters around the main track, and in the CAL the truncated, second-order moments of the energy distribution around the shower axis.
A further selection derives from the different distributions of energy and hits in the ACD between EM and hadron-initiated showers.
At this stage, the hadron rejection power is at the level of 1 to a few $10^{2}$, improving to greater than $10^{3}$ below 100 GeV thanks to the ACD selection.

Similarly to the LAT photon background rejection analysis, the remaining necessary boost in the rejection power is obtained by combining two probability variables that result from training classification trees (CT) to distinguish between EM and hadron events~\cite{LATpaper}.
This is done using large sets of Monte-Carlo (MC) events generated by the accurate LAT simulation package~\cite{LATpaper}, based on the \texttt{Geant4} toolkit~\cite{Geant4}.
Two CTs are used, one built with TKR variables, and a second one based on CAL variables, which describe the complete event topology.
The variables given most weight by the CTs are the same or equivalent to those described above. 
The classifiers allow selection of the electrons through a multitude of parallel paths, each with different selections, that map the many different topologies of the signal events into a single, continuous probability variable that is used to simultaneously handle all valid selections.
The TKR and CAL electron probabilities are finally combined to create an energy-dependent selection that identifies electrons with greater efficiency and optimized background rejection with respect to a single sequence of cuts.
The resulting rejection power is flat and better than $1:10^{3}$ up to 200 GeV and from there rises steadily to $\sim 1:10^{4}$ at 1 TeV in a manner that partially compensates for the increasingly larger relative proton fluxes with energy.
Conversely, the electron selection efficiency, calculated as the ratio of selected versus triggered events, has a peak value of 50\% at 20 GeV and steadily decreases down to 12.5\% at 1 TeV.

\begin{figure}[!t]
\includegraphics[width=\linewidth]{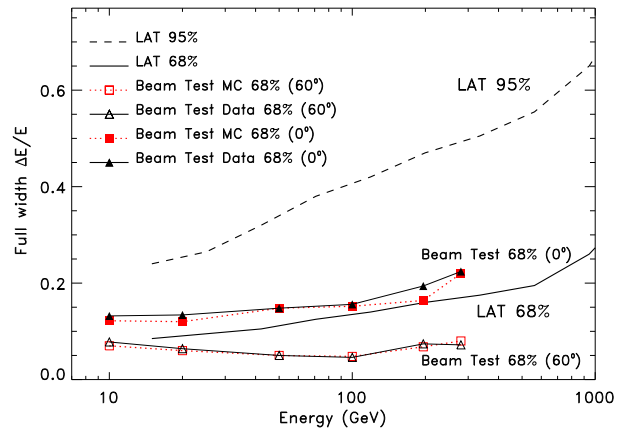}
 \caption{\label{eres}(color online) Energy resolution for the LAT after electron selection; the full widths of the smallest energy window containing the 68\% and the 95\% of the energy dispersion distribution are shown. The comparison with beam test data up to 282 GeV and for on-axis and at 60$^\circ$ incidence shown in the figure indicates good agreement with the resolution estimated from the simulation.}
\end{figure}

{\it Energy reconstruction and validation.} -- Energy reconstruction is the other critical aspect of this analysis.
For EM cascades of several hundreds of GeV a large fraction of the energy falls outside of the LAT CAL. 
The shower imaging capability is therefore crucial in fitting the longitudinal shower profile in order to correct for the energy leakage and estimate the incoming energy with good accuracy.
The resulting energy resolution for events passing the electron selection is shown in figure~\ref{eres}. 
Since showers are not fully contained above 20 GeV, the distribution of the reconstructed energy after leakage correction is asymmetric, with a longer tail toward lower energies.
For this reason we quote the full width of the $68\%$ containment of the distribution as our energy resolution, and check that the full $95\%$ containment does not imply indefinitely long tails; see figure~\ref{eres}.
Candidate electrons traverse on average $12.5$ radiation lengths, resulting from the total thickness of the TKR and CAL detectors and the effect of event selection.

\begin{figure}[!t]
\includegraphics[width=\linewidth]{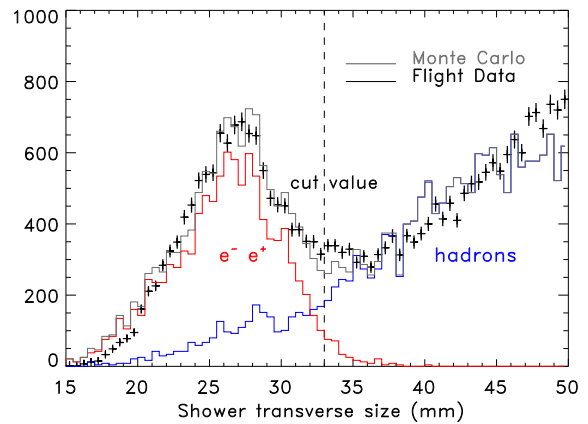}
 \caption{\label{validation}(color online) Distribution of the transverse sizes of the showers (above 150 GeV) in the CAL at an intermediate stage of the selection, where a large contamination from protons is still visible. Flight data (black points) and MC (gray solid line) show very good agreement; the underlying distributions of electron and hadron samples are visible in the left (red) and the right (blue) peaks respectively.}
\end{figure}

The energy reconstruction algorithm and the event analysis rely heavily on the LAT MC simulation.
This was extensively verified and fine-tuned using beam test data for electrons and hadrons up to 282 GeV~\cite{btpaper}.
Extensive efforts are made to avoid bias in the event selection by systematically comparing flight data and MC distributions of likely discriminants of electrons and hadrons, and choosing only those that indicate a good agreement.
Figure~\ref{validation} shows the very good data--MC agreement for the critical variable that maps the transverse shower size.

Systematic uncertainties are determined for all energy bins.
For each step in the event selection, we scan a range of thresholds around the reference value used by the cut and derive the corresponding flux versus GF curve.
We extrapolate the curve to a GF consistent with a null cut, and take the relative difference of the corresponding flux and the reference as the systematic uncertainty associated with the cut.
All such contributions, taken separately with their signs, and the uncertainty of the residual contamination, derived from an overall $20\%$ uncertainty in the underlying proton spectrum are summed in quadrature. 
The result is shown in table~\ref{table:FermiCREdata}.

The final tuning of the event selection provides a maximum systematic error less than $20\%$ at 1 TeV.
The absolute LAT energy scale, at this early stage of the mission, is determined with an uncertainty of $^{+5\%}_{-10 \%}$.
This estimate is being further constrained using flight and beam test data.
The associated systematic error is not folded into those above as it is a single scaling factor over the whole energy range.
Its main effect is to rigidly shift the spectrum by $^{+10\%}_{-20 \%}$ without introducing significant deformations.

While event selection is explicitly energy-dependent to suppress the larger high-energy background, it is not optimized versus the incident angle of incoming particles. 
Nonetheless we have compared the spectra from selected restricted angular bins with the final spectrum reported here; they are consistent within systematic uncertainties.
A further validation of the event selection comes from an independent analysis, developed for lower-energy electrons, which produces the same results when extended up to the the endpoint of its validity at $\sim$ 100 GeV. 
Our capability to reconstruct spectral features was tested using the LAT simulation and the energy response from figure~\ref{eres}.
We superimposed a Gaussian line signal, centered at 450 $\pm$ 50 GeV rms, on a power law spectrum with an index of $3.3$. 
This line contains a number of excess counts as from the ATIC paper~\cite{atic}, rescaled with the LAT GF. 
We verified that this analysis easily detects this feature with high significance (the full width of the 68\% containment energy resolution of the LAT at 450 GeV is 18\%).

\begin{figure}[!t]
\includegraphics[width=\linewidth]{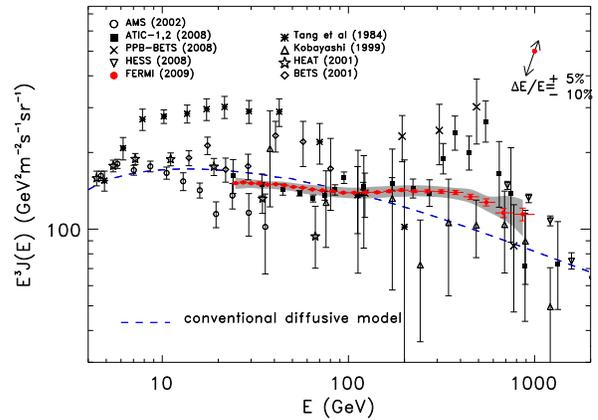}
 \caption{\label{spectrum}
(color) The Fermi LAT CR electron spectrum (red filled circles). 
Systematic errors are shown by the gray band. 
The two-headed arrow in the top-right corner of the figure gives size and direction of the rigid shift of the spectrum implied by a shift of  $^{+5\%}_{-10 \%}$ of the absolute energy, corresponding to the present estimate of the uncertainty of the LAT energy scale. 
Other high-energy measurements and a conventional diffusive model~\cite{SMR2004} are shown.}
\end{figure}

{\it Results and discussion.} -- More than 4M electron events above 20 GeV were selected in survey (sky scanning) mode from 4 August 2008 to 31 January 2009.
Energy bins were chosen to be the full width of the $68\%$ containment of the energy dispersion, evaluated at the bin center.
The residual hadronic background was estimated from the average rate of hadrons that survive electron selection in the simulations, and subtracted from the measured rate of candidate electrons.
The result is corrected for finite energy redistribution with an unfolding analysis~\cite{agostini} and converted into a flux $J_E$ by scaling with the GF, see table~\ref{table:FermiCREdata}.
The distribution of $E^{3} \times J_E$ is shown in table~\ref{table:FermiCREdata} and in figure~\ref{spectrum}.

Fermi data points visually indicate a suggestive deviation from a flat spectrum.
However, if we conservatively add point--to--point systematic errors from table~\ref{table:FermiCREdata} in quadrature with statistical errors, our data are well fit by a simple normalized $E^{-3.04}$ power law ($\chi^{2}=9.7$, d.o.f. 24). 

For comparison, we show a conventional model~\cite{SMR2004} for the electron spectrum, which is
also being used as a reference in a related Fermi-LAT paper~\cite{gevexcess} on the Galactic diffuse gamma-ray emission. 
This uses the GALPROP code~\cite{galprop},
with propagation parameters adjusted to fit a variety of {\it pre-Fermi} CR data, including electrons.
This model has an electron injection spectral index of $2.54$ above 4 GeV, a diffusion coefficient varying with energy as $E^{1/3}$, and includes a diffusive reacceleration term. 
As can be clearly seen from the blue dashed line in figure~\ref{spectrum}, this model produces too steep a spectrum after propagation to be compatible with the Fermi measurement reported here.

\begingroup
\begin{table}[!t]
\scriptsize
\begin{tabular}{ccccc}
\hline
Energy & GF & Residual & Counts & $E^3 \cdot J_E$\\
(GeV) & (${\rm m}^2 {\rm sr}$)& contamination & & (${\rm GeV}^{2}{\rm s}^{-1}{\rm m}^{-2}{\rm sr}^{-1}$)\\
\hline
\hline
23.6--26.0 & 1.65 & 0.04 & 478929 & $151.6 \pm 1.2^{+7.3}_{-8.3}$\\
26.0--28.7 & 2.03 & 0.05 & 502083 & $152.6 \pm 0.9^{+6.2}_{-7.3}$\\
28.7--31.7 & 2.35 & 0.05 & 487890 & $151.4 \pm 0.8^{+5.1}_{-6.5}$\\
31.7--35.0 & 2.59 & 0.09 & 459954 & $151.3 \pm 1.8^{+5.2}_{-6.5}$\\
35.0--38.8 & 2.67 & 0.07 & 385480 & $149.6 \pm 0.7^{+4.4}_{-5.8}$\\
38.8--43.1 & 2.72 & 0.08 & 330061 & $150.2 \pm 0.7^{+4.5}_{-6.0}$\\
43.1--48.0 & 2.76 & 0.10 & 276105 & $148.6 \pm 0.7^{+4.9}_{-6.2}$\\
48.0--53.7 & 2.79 & 0.11 & 233877 & $146.5 \pm 0.7^{+4.9}_{-6.1}$\\
53.7--60.4 & 2.77 & 0.12 & 194062 & $145.5 \pm 0.7^{+5.0}_{-7.1}$\\
60.4--68.2 & 2.76 & 0.13 & 155585 & $143.2 \pm 0.7^{+5.6}_{-6.8}$\\
68.2--77.4 & 2.73 & 0.14 & 126485 & $141.9 \pm 0.8^{+5.6}_{-7.0}$\\
77.4--88.1 & 2.71 & 0.14 & 100663 & $140.8 \pm 0.8^{+6.2}_{-7.0}$\\
88.1--101 & 2.68 & 0.15 & 77713 & $139.0 \pm 0.9^{+6.4}_{-6.8}$\\
101--116 & 2.64 & 0.16 & 61976 & $139.0 \pm 0.9^{+6.4}_{-7.2}$\\
116--133 & 2.58 & 0.17 & 46865 & $139.4 \pm 1.0^{+6.9}_{-7.2}$\\
133--154 & 2.52 & 0.17 & 35105 & $139.5 \pm 1.2^{+7.2}_{-7.4}$\\
154--180 & 2.44 & 0.17 & 27293 & $140.8 \pm 1.3^{+6.9}_{-7.4}$\\
180--210 & 2.36 & 0.18 & 19722 & $142.3 \pm 1.5^{+7.1}_{-7.4}$\\
210--246 & 2.27 & 0.18 & 13919 & $140.9 \pm 1.7^{+7.4}_{-6.8}$\\
246--291 & 2.14 & 0.18 & 10019 & $140.9 \pm 1.9^{+7.5}_{-6.7}$\\
291--346 & 2.04 & 0.18 & 7207 & $140.4 \pm 2.2^{+6.7}_{-7.0}$\\
346--415 & 1.88 & 0.18 & 4843 & $139.4 \pm 2.6^{+7.0}_{-7.2}$\\
415--503 & 1.73 & 0.19 & 3036 & $134.0 \pm 3.1^{+9.3}_{-7.5}$\\
503--615 & 1.54 & 0.20 & 1839 & $127.4 \pm 4.1^{+8.7}_{-8.6}$\\
615--772 & 1.26 & 0.21 & 1039 & $115.8 \pm 4.8^{+15.2}_{-10.9}$\\
772--1000 & 0.88 & 0.21 & 544 & $114.4 \pm 6.5^{+19.1}_{-17.8}$\\ 
\hline
\end{tabular}
\caption{\label{spectrumtable}
Geometry factor, residual contamination, number of counts before background subtraction and the flux $J_E$ multiplied by $E^{3}$. 
The statistical error is followed by the systematic error. 
The latter does not include the effect due to the uncertainty in the absolute energy scale.}

\label{table:FermiCREdata}
\end{table}
\endgroup

The observation that the spectrum is much harder than the conventional one may be explained by assuming a harder electron spectrum at the source, which is not excluded by other measurements. 
However, the significant flattening of the LAT data above the model predictions for E $\geq 70$ GeV may also suggest  the presence of one or more local sources of high energy CR electrons. 
We found that the LAT spectrum can be nicely fit by adding an additional component of primary electrons and positrons, with injection spectrum $J_{\rm extra}(E) \propto E^{- \gamma_{\rm e}} \exp\{-E/E_{\rm cut}\}$, $E_{\rm cut}$ being the cutoff energy of the source spectrum. 
The main purpose of adding such a component is to reconcile theoretical predictions with both the Fermi electron data and the Pamela data~\cite{pamela} showing an increase in the $e^{+}/(e^{-} + e^{+})$ fraction above 10 GeV. 
The latter cannot be produced by secondary positrons coming from interaction of the Galactic CR with the ISM. 
Such an additional component also provides a natural explanation of the steepening of the spectrum above 1 TeV indicated by H.E.S.S. data~\cite{hess}. 
As discussed in~\cite{Profumo:2008ms} and references therein, pulsars are the most natural candidates for such sources. 
Other astrophysical interpretations (e.g.~\cite{blasi:0903}), or dark matter scenarios, can not be excluded at the present stage.

A detailed discussion of theoretical models lies outside the scope of this work and will be presented in a forthcoming Fermi interpretation paper. 
At this stage it suffices to say that Fermi will significantly change the understanding of this part of the electron spectrum. 

The $Fermi$ LAT Collaboration acknowledges generous ongoing support from a number of agencies and institutes that have supported both the development and the operation of the LAT as well as scientific data analysis.  These include the National Aeronautics and Space Administration and the Department of Energy in the United States, the Commissariat \`a l'Energie Atomique and the Centre National de la Recherche Scientifique / Institut National de Physique Nucl\'eaire et de Physique des Particules in France, the Agenzia Spaziale Italiana and the Istituto Nazionale di Fisica Nucleare in Italy, the Ministry of Education, Culture, Sports, Science and Technology (MEXT), High Energy Accelerator Research Organization (KEK) and Japan Aerospace Exploration Agency (JAXA) in Japan, and the K.~A.~Wallenberg Foundation, the Swedish Research Council and the Swedish National Space Board in Sweden.

Additional support for science analysis during the operations phase from the following agencies is also gratefully acknowledged: the Istituto Nazionale di Astrofisica in Italy and the K.~A.~Wallenberg Foundation in Sweden for providing a grant in support of a Royal Swedish Academy of Sciences Research fellowship for JC.


\begin{thebibliography}{}\label{sec:TeXbooks}

\bibitem{SMR2004}
A.~W.~Strong, I.~V.~Moskalenko and O.~Reimer, \jref{\japj}{613}{962}{2004}.

\bibitem{kobayashi}
J.~Nishimura \etal, \jref{\japj}{238}{394}{1980};
J.~Nishimura \etal, \jref{\jasr}{19}{767}{1997};
T.~Kobayashi \etal, \jref{\japj}{601}{340}{2004}.

\bibitem{ams1}
M.~Aguilar \etal, \jref{\jprep}{366}{331}{2002}.

\bibitem{galprop}
I.~Moskalenko and A.~Strong, \jref{\jasr}{27}{717}{2001};
http://galprop.stanford.edu.

\bibitem{aharonian}
F.~Aharonian, A.~Atoyan and H.~J.~Voelk, \jref{\jaa}{294}{L41}{1995}.

\bibitem{pohl}
M.~Pohl and J.~A.~Esposito, \jref{\japj}{507}{327}{1998}.

\bibitem{pamela}
O.~Adriani \etal, \jarXiv{0810.4995}.

\bibitem{atic}
J.~Chang \etal, \jref{\jnat}{456}{362}{2008}.

\bibitem{hess}
F.~Aharonian \etal, \jref{\jprl}{101}{261104}{2008}.

\bibitem{ppbbets}
S.~Torii \etal, \jarXiv{0809.0760};
\jref{\japj}{559}{973}{2001}.

\bibitem{shen}
C.~S.~Shen, \jref{\japj}{162}{L181}{1970}.

\bibitem{Profumo:2008ms}
S.~Profumo, \jarXiv{0812.4457}.

\bibitem{cholis}
I.~Cholis \etal, \jarXiv{0811.3641}; 
\jarXiv{0810.5344}.

\bibitem{LATpaper}
W.~Atwood \etal, \japj\ \jtbp;
\jarXiv{0902.1089}.

\bibitem{durban}
J.~F.~Ormes {\it et al.}, \jproc{International Cosmic Ray Conference}{Durban}{1997}, edited by M.~S.~Potgieter, C.~Raubenheimer, and D.~J. van der Walt, Transvaal, South Africa Potchefstroom University.

\bibitem{merida}
A.~Moiseev, {\it et al.}, \jproc{International Cosmic Ray Conference}{Merida}{2007}, edited by R.~Caballero, J.~C.~D'Olivo, G.~Medina-Tanco, L.~Nellen, F.~A.~Sánchez, J.~F.~Valdés-Galicia, Universidad Nacional Autónoma de México, Mexico City, Mexico, Vol.2, 449,
\jarXiv{0706.0882}.

\bibitem{egretdiffuse}
P.~Sreekumar {\it et al.}, \jref{\japj}{494}{523}{1998}.

\bibitem{Geant4}
S.~Agostinelli \etal, \jref{\jnima}{506}{250}{2003}.

\bibitem{btpaper}
L.~Baldini \etal, \jproc{First International GLAST Symposium}{Stanford University}{2007}, edited by S.~Ritz, P.~Michelson, C.~Meegan, AIP Conference Proceedings Vol.921, 190.

\bibitem{agostini}
G.~D'Agostini, \jref{\jnima}{362}{487}{1995}.

\bibitem{gevexcess}
A.~A.~Abdo \etal, in preparation.

\bibitem{blasi:0903}
P.~Blasi, \jarXiv{0903.2794v1}.

\end{thebibliography}
\end{document}